\documentclass[fleqn,usenatbib]{mnras}

\usepackage{newtxtext,newtxmath}

\usepackage[T1]{fontenc}
\usepackage{multirow}
\DeclareRobustCommand{\VAN}[3]{#2}
\let\VANthebibliography\thebibliography
\def\thebibliography{\DeclareRobustCommand{\VAN}[3]{##3}\VANthebibliography}

\usepackage{graphicx}	
\usepackage{amsmath}	
\usepackage{amssymb}	

\let\oldAA\AA
\renewcommand{\AA}{\text{\normalfont\oldAA}}

\title[Two new $z>6$ RL QSOs]{New Radio-Loud QSOs at the end of the Re-ionisation Epoch}

\author[L. Ighina et al.]{
L. Ighina$^{1,2,3}$\thanks{E-mail: lighina@uninsubria.it},
A. Caccianiga$^{1}$,
A. Moretti$^{1}$,
S. Belladitta$^{1,2}$,
J. W. Broderick$^{3}$,
G. Drouart$^{3}$,
\newauthor{ J. K. Leung$^{4,5,6}$, N. Seymour$^{3}$}
\\
$^{1}$INAF, Osservatorio Astronomico di Brera, via Brera 28, 20121, Milano, Italy \\
$^{2}$DiSAT, Universit\`a degli Studi dell'Insubria, via Valleggio 11, 22100 Como, Italy\\
$^{3}$ International Centre for Radio Astronomy Research, Curtin University, 1 Turner Avenue, Bentley, WA, 6102, Australia\\
$^{4}$ Sydney Institute for Astronomy, School of Physics, University of Sydney, NSW 2006, Australia\\
$^{5}$ CSIRO Space and Astronomy, PO Box 76, Epping, NSW, 1710, Australia\\
$^{6}$ ARC Centre of Excellence for Gravitational Wave Discovery (OzGrav), Hawthorn, VIC 3122, Australia
}

\date{Accepted XXX. Received YYY; in original form ZZZ}

\pubyear{2015}

\begin{document}
\label{firstpage}
\pagerange{\pageref{firstpage}--\pageref{lastpage}}
\maketitle

\begin{abstract}
We present the selection of high-redshift ($z\gtrsim5.7$) radio-loud (RL) quasi-stellar object (QSO) candidates from the combination of the radio Rapid ASKAP Continuum Survey (RACS; at 888~MHz) and the optical/near-infrared Dark Energy Survey (DES). In particular, we selected six candidates brighter than $S_{\rm 888MHz}>1$~mJy~beam$^{-1}$ and ${\rm mag}(z_\mathrm{{DES}})<21.3$ using the dropout technique (in the $i$-band). From this sample, we were able to confirm the high-$z$ nature ($z\sim6.1$) of two sources, which are now among the highest-redshift RL QSOs currently known. Based on our Gemini-South/GMOS observations, neither object shows a prominent Ly$\alpha$ emission line. This suggests that both sources are likely to be weak emission-line QSOs hosting radio jets and would therefore further strengthen the potential increase of the fraction of weak emission-line QSOs recently found in the literature. However, further multiwavelength observations are needed to constrain the properties of these QSOs and of their relativistic jets. From the discovery of these two sources, we estimated the space density of RL QSOs in the redshift range $5.9<z<6.4$ to be 0.13$^{+0.18}_{-0.09}$ and found it to be consistent with the expectations based on our current knowledge of the blazar population up to $z\sim5$.
 \end{abstract}




\section{Introduction}

The recent advent of wide-area optical and near-infrared (NIR) surveys, such as the Panoramic Survey Telescope and Rapid Response System \citep[Pan-STARRS;][]{Chambers2016} and the Dark Energy Survey \citep[DES;][]{Abbott2018}, has led to the discovery of several hundreds of $z>6$ quasi-stellar objects (QSOs) \citep[e.g.][]{Morganson2012,Banados2014,Banados2016,Jiang2016,Reed2017,Reed2019,Wang2019,Wang2021,Yang2020}. 
The systematic study of these high-$z$ systems in the optical and infrared (IR) has enabled the characterisation of their super massive black hole (SMBH; \citealt{Mazzucchelli2017,Shen2019,Farina2022}), their host galaxy \citep[e.g.][]{Venemans2020, Decarli2022} and their environment \cite[e.g.][]{Balmaverde2017, Ota2018, Chen2022}. However, the properties of their relativistic jets are still poorly constrained. This is mainly due to the fact that the number of high-$z$ QSOs detected in the radio band is still small (see e.g. \citealt{Liu2021}) and, as a consequence, the number of high-redshift radio-loud (RL, i.e. hosting powerful relativistic jets\footnote{Following the literature, we define a source to be RL if its rest-frame radio loudness $R=S_{\rm 5 GHz}$/${S_{\rm 4400 \AA}}>10$. Even though it has been shown that this classification might be too simplistic for the definition of QSOs with relativistic jets, it should include all the most powerful radio jets \citep[e.g.,][]{Padovani2017}.}) QSOs currently known is even smaller. Indeed, only $\sim$10 QSOs are currently known to be RL at $z>6$. While five of them have been discovered in different wide-area radio surveys (J0309+2717 at $z = 6.10$, \citealt{Belladitta2020,Belladitta2022}; J1427+3312 at $z = 6.12$, \citealt{McGreer2006}; J1429+5447 at $z = 6.18$, \citealt{Willot2010}; J2318$-$3113 at $z=6.44$, \citealt{Decarli2018, Ighina2021}; J172.3556+18.7734 at $z=6.82$, \citealt{Banados2021}) the recent release of the LOFAR Two-metre Sky Survey \citep[LoTSS, at 144~MHz;][]{Shimwell2017,Shimwell2019,Shimwell2022}, allowed the discovery of several new RL QSOs at $z>6$ \citep[between 3 and 5, depending on the assumed radio spectral indices;][]{Gloudemans2021,Gloudemans2022}.
Similarly, with the upcoming large-area radio surveys planned with the square kilometre array (SKA\footnote{\url{https://www.skatelescope.org/wp-content/uploads/2011/03/SKA-Astophysics-Vol1.pdf}}; \citealt{Braun2019}) and its precursors, which will reach sensitivities of tens of \textmu Jy at $\sim$1~GHz \citep[e.g.][]{Norris2011,Norris2021}, many new RL active galactic nuclei (AGNs), both QSOs and radio galaxies, are expected to be found at $z\gtrsim6.5$. For example, \cite{Endsley2022,Endsley2022a} very recently discovered the most distant radio galaxy currently known (at $z=6.85$) in a 1.5~deg$^2$ region, which implies the existence of many more similar systems over the entire sky. These new surveys will allow us to build complete samples of RL AGNs, from which to derive conclusive statistical properties on the entire population, well beyond the current redshift limit \citep[$z\sim5.5$, e.g.][]{Caccianiga2019}.
Having a large sample of RL AGNs at high-$z$ will allow for a series of important scientific results, such as constraining the cosmic evolution of SMBHs hosted in systems with powerful jets directly after their formation \citep[e.g.][]{Volonteri2011,Diana2022} and studying the intergalactic medium (IGM) during the epoch of re-ionisation trough neutral hydrogen (HI) absorption at low radio frequencies  \citep[e.g.][]{Carilli2002,Vrbanec2020}.

In this work we present a sample of six $z\gtrsim5.7$ RL QSO candidates built from the the cross-match of the second data release of the DES in the optical/NIR and the first data release of the low-band Rapid ASKAP Continuum Survey \citep[RACS;][]{McConnell2020, Hale2021}, the first large-area radio survey performed with the Australian Square Kilometre Array Pathfinder (ASKAP; \citealt{Hotan2021}). From this sample, we have already identified two new RL QSOs at $z>6$, DES~J032021.431$-$352104.11 ($z=6.13$; DES~J0320$-$35 hereafter) and DES~J032214.541$-$184118.15 ($z=6.09$; DES~J0322$-$18 hereafter).

This work is structured as follows: in Sec. \ref{sec:selection} we outline the optical/NIR and radio criteria adopted for the selection of high-$z$ RL QSO candidates; in Sec. \ref{Sec:two_confirmed} we present the spectroscopic confirmation of the two most promising candidates and their archival optical/NIR and radio data; in Sec. \ref{sec:expectations} we discuss the number of RL QSOs expected to be found at high redshift with future surveys; in Sec. \ref{Sec:summary} we summarise the conclusions of the work.\\ 

Throughout the paper we assume a flat $\Lambda$CDM cosmology with $H_{0}$=70 km s$^{-1}$ Mpc$^{-1}$, $\Omega_m$=0.3 and $\Omega_{\Lambda}$=0.7. Spectral indices are given assuming S$_{\nu}\propto \nu^{-\alpha}$ and all errors are reported at a 68\% confidence level, unless otherwise specified.

\section{Selection Criteria}
\label{sec:selection}
In order to efficiently select high-$z$ RL QSOs, we started from the combination of the second data release of the DES and the first data release of the low-band RACS (simply RACS hereafter) survey and adopted the Lyman-break dropout technique (see e.g. \citealt{Banados2016,Wang2017,Wang2019, Belladitta2019, Belladitta2020, Belladitta2022b}). 


\subsection{Optical and near/mid-infrared criteria}
\begin{figure*}
	\includegraphics[width=0.464\hsize]{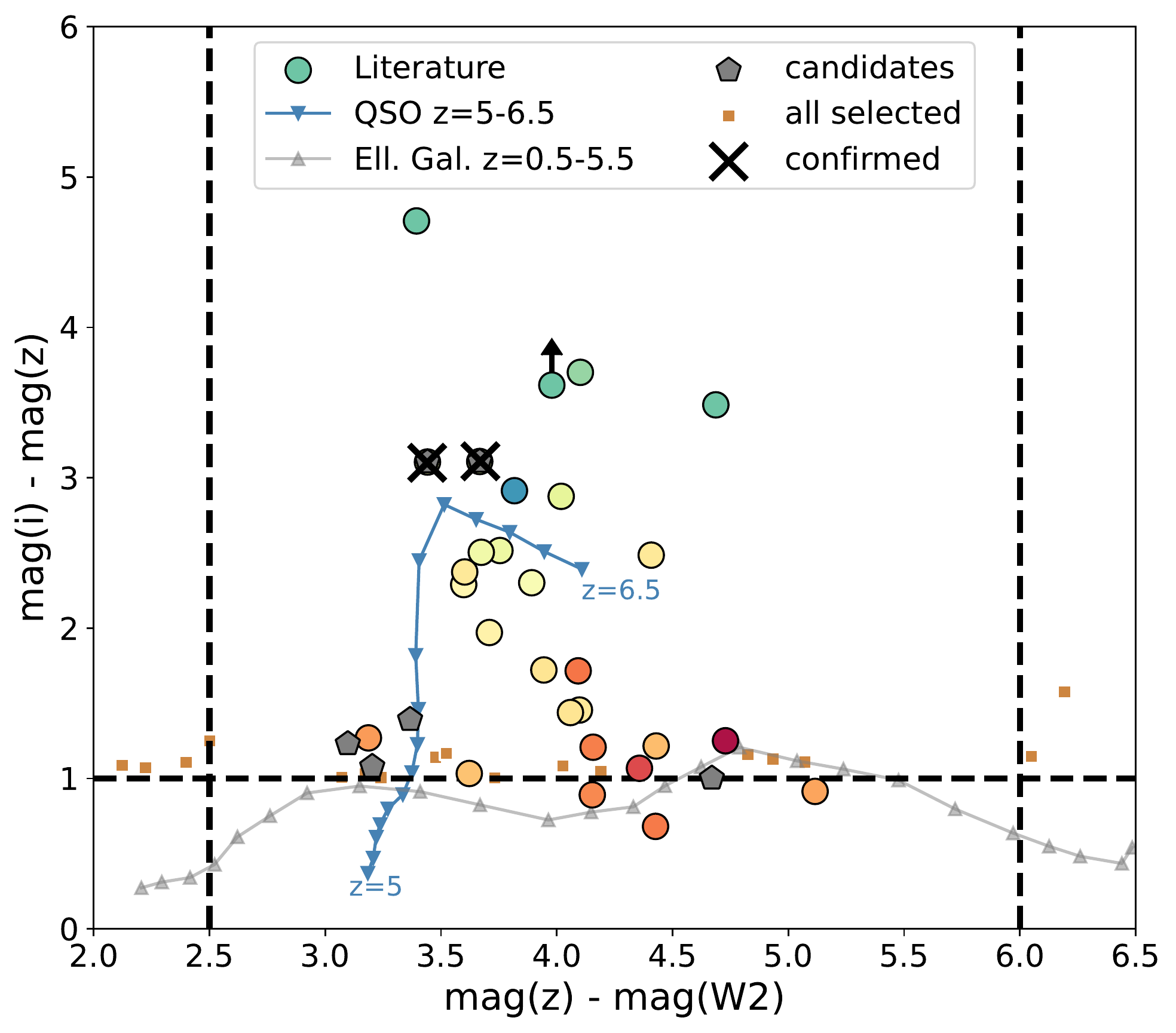}
    \includegraphics[width=0.522\hsize]{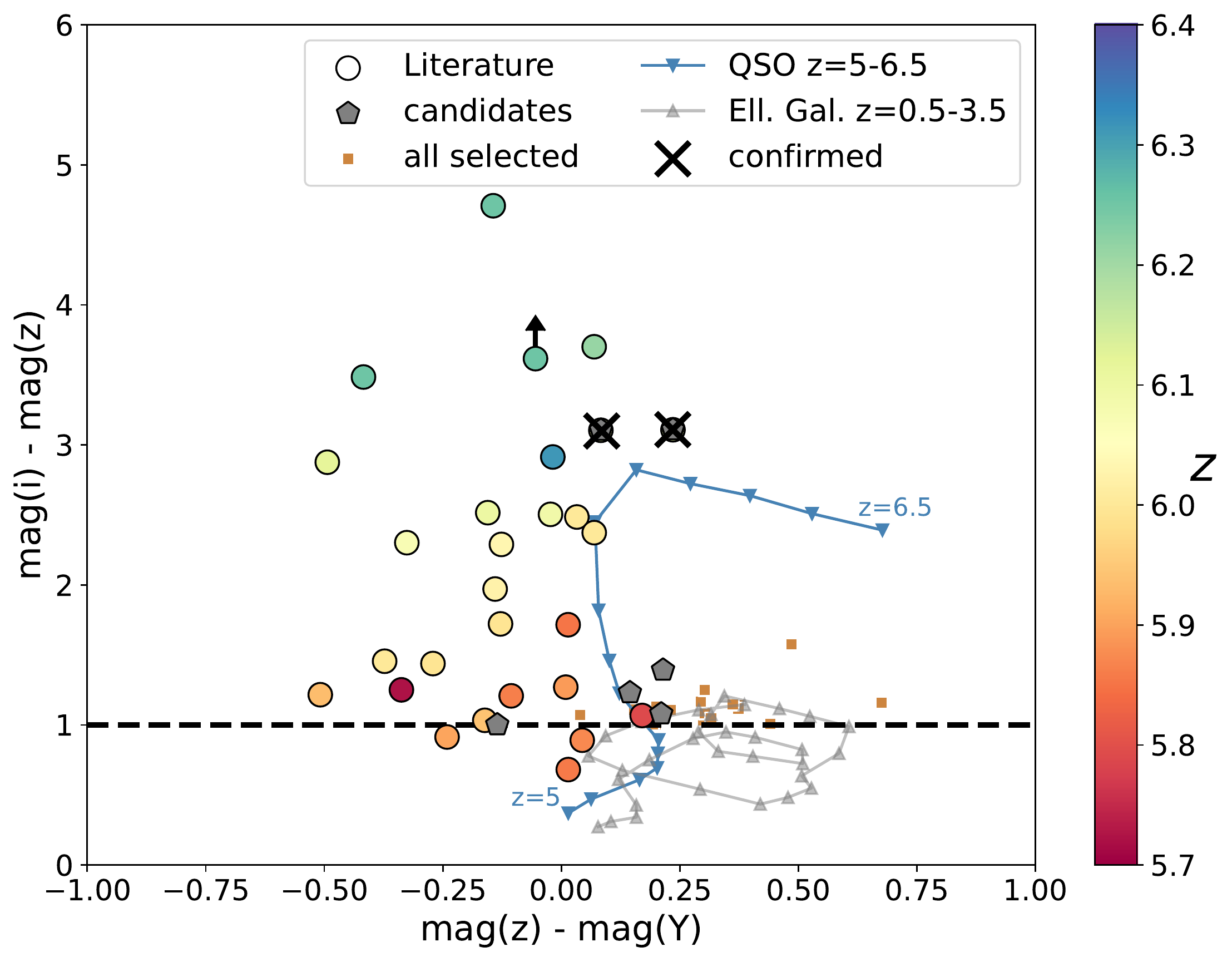}

    \caption{mag($i$) - mag($z$) as a function of the mag($z$) -mag($W2$)  (AB and Vega system respectively; left panel) and as a function of the mag($z$) -mag($Y$) one (right panel) for the $z>5.7$ QSOs currently known in the DES area (colour scale based on their redshift) together with the candidates selected with the criteria described in the text (grey pentagons). The brown squares indicate all the sources selected from the DES catalogue (i.e., not including the catWISE data and before the inspection of the $g$-band images). The blue line represents the expected colours of a typical QSO between redshift $5-6.5$ and with a weak Ly$\alpha$ emission line (given by the combination of the composite spectrum derived in \citealt{Banados2016}, see right panel of Fig. \ref{fig:opt_images} for example, and the QSO template described in \citealt{Polletta2007}). Similarly, the grey line represents the expected colours for a low-redshift ($z=0-3.5$) elliptical galaxy (based on the template of \citealt{Polletta2007}). The limits used for the selection of bona-fide high-$z$ candidates are shown with the horizontal and vertical black dashed lines.}
    \label{fig:colour_cuts}

\end{figure*}

DES covers $\sim$5000~deg$^2$ mostly in the southern extragalactic sky and it is one of the deepest wide-area optical/NIR surveys currently available. The limiting magnitudes at a signal-to-noise (S/N) level of 10 in its second data release are: $g_\mathrm{DES}$=24.7, $r_\mathrm{DES}$=24.4, $i_\mathrm{DES}$=23.8, $z_\mathrm{DES}$=23.1 and $Y_\mathrm{DES}$=21.7 mag \citep{Abbott2021}.
In order to select bona-fide high-$z$ QSO candidates, we accessed the products of DES through the dedicated SQL portal\footnote{Available at: \url{https://des.ncsa.illinois.edu/desaccess/home}.}.
To be as complete as possible, we set our criteria to recover almost all the $5.7<z<6.4$ QSOs (both radio loud and radio quiet) with a magnitude in the $z_{\rm DES}$-band $<21.3$ already discovered in the DES area (28 in total), even if they were originally selected from another survey (i.e. with different filter sets). We adopted the following criteria for the candidate selection:

\begin{itemize}

    \item \texttt{mag\_auto\_z} $<21.3$,
    \item \texttt{class\_star\_z} $>0.85$,
    \item \texttt{magerr\_auto\_z} $<0.3$ and  \texttt{magerr\_auto\_y} $<0.3$,
    \item \texttt{mag\_auto\_i} $-$ \texttt{mag\_auto\_z} $>1$,
    \item non detection in the $g-$band\footnote{To assure a reliable non-detection in this band we firstly considered only candidates with \texttt{mag\_auto\_g} $>23.5$ and \texttt{magerr\_auto\_g} $>0.3$. Then, after having applied all the other selection criteria, we checked the single $g$-band images and discarded all the sources  where a  S/N $\gtrsim3$ signal was present.},

\end{itemize}
where all the magnitudes are in the AB system. These criteria are meant to select primarily compact stellar-like objects with a very faint emission in the $g$-band and an $i-z$ dropout due to the Ly$\alpha$ absorption at high redshift and not to an artefact in the images. The total number of sources selected with these optical/NIR criteria (without the inspection of the $g$-band images) and with a radio association (see next subsection) is 25.
After filtering the DES catalogue with the criteria reported above, we also considered mid-IR data from the catalogue {\it Wide-field Infrared Survey Explorer} \citep[catWISE;][]{Eisenhardt2020}. In particular, by using a cross-match radius of 2\arcsec, we only kept candidates with $2.5 < \texttt{mag\_auto\_z} - \texttt{mag\_W2} < 6$ (where the WISE magnitude is in the Vega system). This criterion should help us avoid absorbed QSOs and a subset of elliptical galaxies at low redshift \citep[e.g.][]{Carnall2015}. In Fig. \ref{fig:colour_cuts} we show the $i-z$ vs $z-W2$ (left panel) and the $i-z$ vs $z-Y$ (right panel) colours for the $z>5.7$ QSOs currently known in the DES area as well as for the selected candidates. 
Based on the colour cuts applied above, we recovered all but one of the already known QSOs at $z>5.9$ (1 out of 20 sources not included, due to the \texttt{class\_star\_z} criterion). Whereas, three $5.7<z<5.9$ QSOs were missed due to their small $i-z$ colour value (3 out of 8, see Fig. \ref{fig:colour_cuts}). This is because at these redshifts, the Ly$\alpha$ emission line ($\sim$8150--8400~\AA \, in the observed frame) and its dropout shift to wavelengths covered by the $i$-band filter ($\sim$7100--8500~\AA). A systematic selection focused on the specific $5<z<5.9$ redshift range will be presented in a future work.

\subsection{Radio association}
In order to only select radio-bright candidates, we cross-matched the optically-selected sources with the first data release of RACS. This survey covers the entire sky south of +41$^{\rm o}$ in declination with a median RMS of $\sim$0.25~mJy~beam$^{-1}$ and is also expected to perform more scans of the same area in the future also at higher frequencies (RACS-mid at $\sim$1.36~GHz and RACS-high at $\sim$1.65~GHz). We restricted our selection only to objects with a peak flux density $S_\mathrm{peak}$\textgreater1~mJy~beam$^{-1}$ and with a $S_\mathrm{int}$/$S_\mathrm{peak}$\textless 1.5 in order to select unresolved sources only, and therefore, reducing the position uncertainty of the radio counterpart. This last criterion is satisfied by all the 13 $z>5$ RL QSOs already known in RACS area. Given the typical positional errors of RACS (about 2--3\arcsec, see section 3.4.3 in \citealt{McConnell2020}), we adopted a search radius of 3\arcsec \, between the optical and the radio counterparts. We note that for this first search, we made use of the source lists derived from the images described in \cite{McConnell2020}\footnote{Available from the CASDA website: \url{https://data.csiro.au/domain/casdaObservation?redirected=true}.} since the final catalogue of RACS-low survey built by \cite{Hale2021}, where all the images have been combined together after being convolved to a common resolution of 25\arcsec, would further increase the position uncertainty, especially for faint sources. Nevertheless, in the following, to ensure a reliable flux density measurement when presenting a radio flux density of a source from RACS, we will consider the values reported in the \cite{Hale2021} catalogue.

We also cross-matched the optically selected candidates with the catalogue derived from the `convolved' images of the pilot evolutionary map of the Universe (EMU) survey \citep{Norris2021}, which is at a similar frequency (944~MHz), it has a similar angular resolution ($\sim$18\arcsec), but it is significantly deeper (RMS $\sim 25-30$~\textmu Jy~beam$^{-1}$ over $\sim$270~deg$^2$). By adopting a radio-optical association of 1.5\arcsec \, (which is enough to account for both the accuracy and precision uncertainties of the sources in the catalogue, see \citealt{Norris2021}) and the same optical/NIR and radio criteria reported above, we select one more $>1$~mJy~beam$^{-1}$ candidate.

Finally, for sources at dec $>-40^{\rm o}$ and bright enough to be detected in the VLA Sky Survey (VLASS; resolution $\sim2.5''$ at 3~GHz; \citealt{Lacy2020}), we also checked their radio images in this survey and only kept the candidates if a $>$3$\times$RMS signal is present within 1.2\arcsec \, from the optical/NIR counterpart. In this way we discarded 3 sources out of the 11 covered by VLASS.

The total number of candidates selected with the optical/IR and radio criteria described above is six and we report in Tab. \ref{tab:list_candiates} their main properties used for the selection. In the following we focus on two of them, which have already been confirmed spectroscopically, while the remaining candidates will be part of a future publication.

\begin{table*}
	\centering
	\begin{tabular}{lccccccccc}
	NAME & RA & Dec  & $S_{\rm 888MHz}$ & $z_\mathrm{DES}$ & $(i-z)_{\rm DES}$ & $z_\mathrm{DES}-W2$ & distance$_{\rm ro}$ & radio survey\\
	 & (deg) & (deg)  & (mJy~beam$^{-1}$) & (mag AB) &  & (mag AB -- Vega) & (arcsec) &\\
\hline
    DES J0249$-$28 & 42.311270 & $-$28.855045  & 5.06$\pm$0.27 & 20.83$\pm$0.03 & 1.23 & 3.09 & 2.01 & RACS\\
	DES J0320$-$35 $\checkmark$& 50.089335 & $-$35.351148 & 3.21$\pm$0.28 & 20.59$\pm$0.03 & 3.11 & 3.28 & 1.32 & RACS\\
	DES J0322$-$18 $\checkmark$& 50.560622 & $-$18.688189 & 1.64$\pm$0.19 & 20.94$\pm$0.04 & 3.11 & 3.44 & 2.59 & RACS\\
    DES J0427$-$53 & 66.930894 & $-$53.39384 & 3.61$\pm$0.33 & 20.04$\pm$0.01 & 1.08 & 3.20 & 1.47 & RACS\\
    DES J0616$-$48 & 94.199469 & $-$48.347094  & 2.13$\pm$0.42 & 21.24$\pm$0.03 & 1.39 & 3.37 & 2.76 & RACS\\
    DES J2020$-$62 & 305.170194 & $-$62.252554  & 1.09$\pm$0.06* & 19.23$\pm$0.01 & 1.00 & 4.67 & 0.80 & EMU \\

	\hline
	
	\end{tabular}

\caption{{\bf Col. (1, 2, 3):} name and optical coordinates of the source. The check-mark beside the name indicate the sources that have already been confirmed; {\bf Col. (4):} peak flux densities (with the corresponding error) at 888~MHz from the RACS catalogue derived by {\protect\cite{Hale2021}} (* indicates flux densities at 944~MHz from EMU, whose uncertainty corresponds to the RMS of the image); {\bf Col. (5):} $z$-band magnitude from DES with the corresponding error; {\bf Col. (6):} dropout value ($i-z$); {\bf Col. (7):}  $z_\mathrm{DES}-W2$ colour, where $z_\mathrm{DES}$ is in the AB system and $W2$ in the Vega system; {\bf Col. (8):} radio-to-optical distance; {\bf Col. (9):} radio survey from which the candidate was selected.}  	
	\label{tab:list_candiates}
\end{table*}

\subsection{Completeness and contamination}
With the optical/IR thresholds reported above, we recover about 95 per cent of the already discovered $z>5.9$ QSOs, while the remaining 5 per cent (one object) does not satisfy the point-like criterion. We note that this value does not take into account the actual completeness of the literature sources in the DES field. Several works showed that the high-z QSO selections based on colour cuts (such as ours) can be considered almost complete for relatively bright high-$z$ QSOs (magnitudes $<22$), see, for example,
fig. 3 in \cite{Willott2010b} and fig. 10 in \cite{Kim2022}. Thanks to the radio counterpart requirement, our selection criteria are less strict compared to the usual cuts adopted in other works based on the optical/IR colours only, both in terms of the dropout value and in the colours of the continuum (i.e. the $z-Y$ colour; see e.g. \citealt{Venemans2013,Banados2016,Reed2017,Reed2019}). Therefore, in the following, we adopt the 95 per cent value as the best estimate of the completeness of our selection, noting that only significantly different values (completeness $<40$ per cent) would change our results in sec. \ref{sec:expectations}, given the limited statistics.

\begin{figure}
	\includegraphics[width=\hsize]{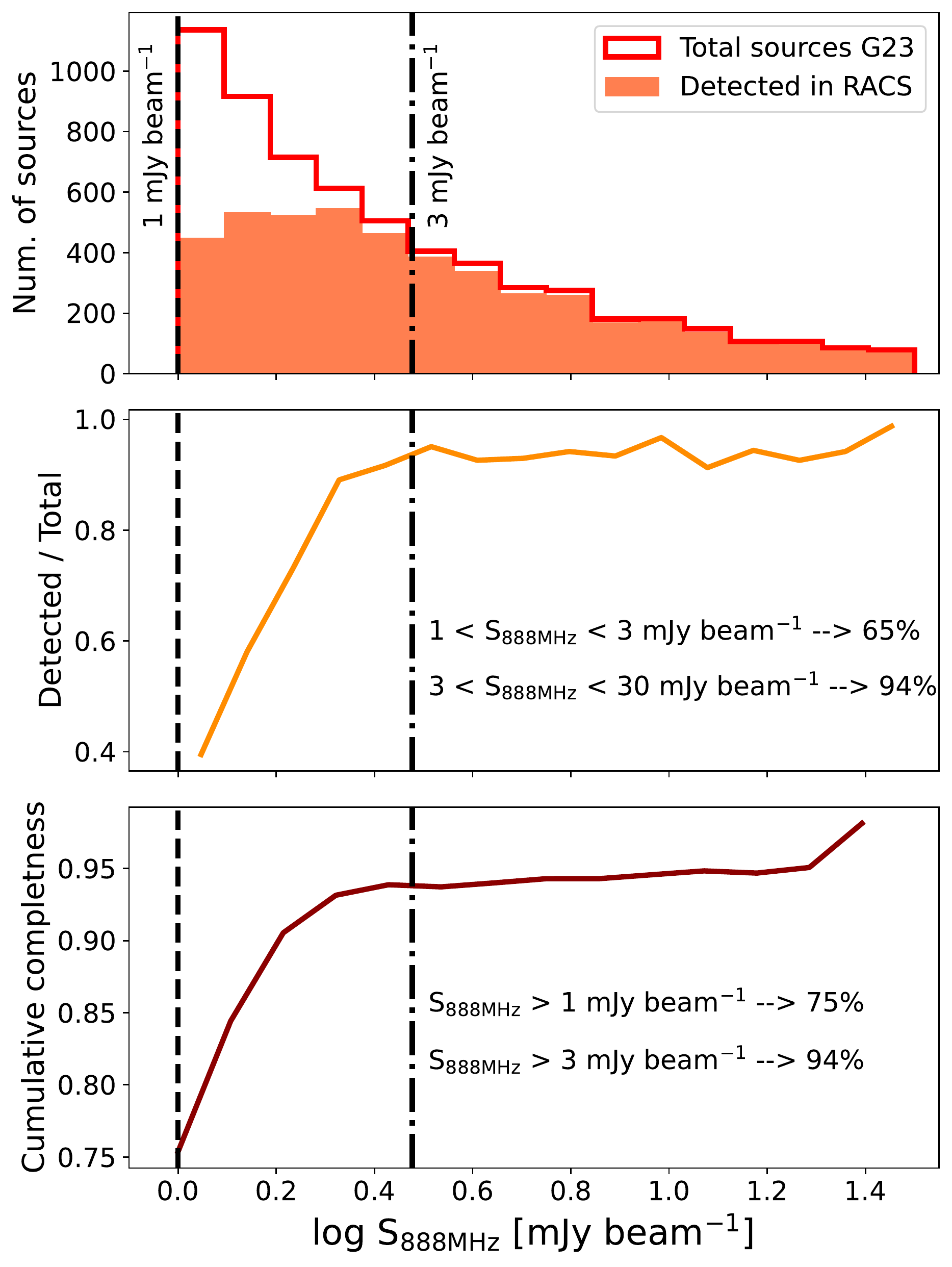}

\caption{Fraction of point-like radio sources present in the G23 field with $1<$~S$_\mathrm{888MHz}<30$~mJy~beam$^{-1}$ detected in RACS. {\bf Top:} flux density distribution of all the $1<$~S$_\mathrm{888MHz}<30$~mJy~beam$^{-1}$ point-like sources present in the G23 field (empty histogram), together with the distribution of the ones detected in RACS (filled histogram); {\bf Centre:} detection fraction in RACS per flux density bin; {\bf Bottom:} cumulative fraction of sources detected in RACS as a function of flux density.}
    \label{fig:completness}
\end{figure}

In order to estimate the completeness related to the radio criteria adopted for RACS, we considered the ASKAP observations of the 23$^{\rm rd}$ GAlaxy Mass Assembly (G23; \citealt{Driver2011}) field described in \cite{Gurkan2022}. This field covers 83~deg$^2$ in the southern hemisphere (centred at RA = 23h and Dec = $-$32$^{\rm o}$) and was observed in the radio band with the ASKAP telescope as part of the EMU Early Science programme. These observations are similar to RACS both in terms of central frequency and resolution (888~MHz and $\sim$10\arcsec, respectively), but they are about an order of magnitude deeper (RMS$\sim$38~\textmu Jy~beam$^{-1}$). From the catalogue derived by \cite{Gurkan2022}, we considered all the sources with S$_\mathrm{int}$/S$_\mathrm{peak} < 1.5$, to ensure the selection of point-like objects only, and with $1<$~S$_\mathrm{peak}<30$~mJy~beam$^{-1}$, that is, with radio flux densities similar to our selected high-$z$ QSO candidates, and evaluated the fraction of these objects detected in RACS. In the top panel of Fig. \ref{fig:completness} we report the flux density distribution of all the point-like $1<$~S$_\mathrm{peak}<30$~mJy~beam$^{-1}$ G23 objects (empty histogram) and the subset detected in the RACS catalogue (filled histogram).
In the lower panels we report the fraction of sources detected in RACS per bin of flux density (central panel) and their cumulative completeness fraction above a given flux density (bottom panel). Based on the detection fraction in RACS, we expect our overall radio selection to be about 75 per cent complete for S$_\mathrm{888MHz}>1$~mJy~beam$^{-1}$. In particular, while for relatively large radio flux densities (S$_\mathrm{888MHz}>3$~mJy~beam$^{-1}$) the detection fraction is almost one ($\sim$94 per cent) the detection fraction drops to about 65 per cent for faint radio sources ($1<$~S$_\mathrm{888MHz}<3$~mJy~beam$^{-1}$). These values are similar to the ones derived by \cite{Hale2021} for the RACS catalogue after combining the different images to a common resolution (see their fig. 15, left panel).\\
Based on these values, and the total number of candidates selected from RACS (see Tab \ref{tab:list_candiates}), in our selection we expect to be missing about two candidates ($\sim$2.04). Moreover, if we consider the overall area of the DES ($\sim$5000~deg$^2$), we would expect to find $\sim$0.4 sources in the region covered by the EMU pilot survey ($\sim$270~deg$^2$), which is consistent with the selection of one candidate in this area.

Finally, in order to determine the number of spurious associations we can expect from our selection, we cross-matched the DES catalogue with the RACS and EMU ones by applying a rigid relative shift in their positions and then applied the same selection criteria used for the selection of our candidates. In this way, we found that the expected number of spurious associations is $1.6\pm0.5$.

\section{Discovery of two new $\lowercase{z}>6$ RL QSO\lowercase{s}}
\label{Sec:two_confirmed}
\begin{figure*}
	\includegraphics[width=0.37\hsize]{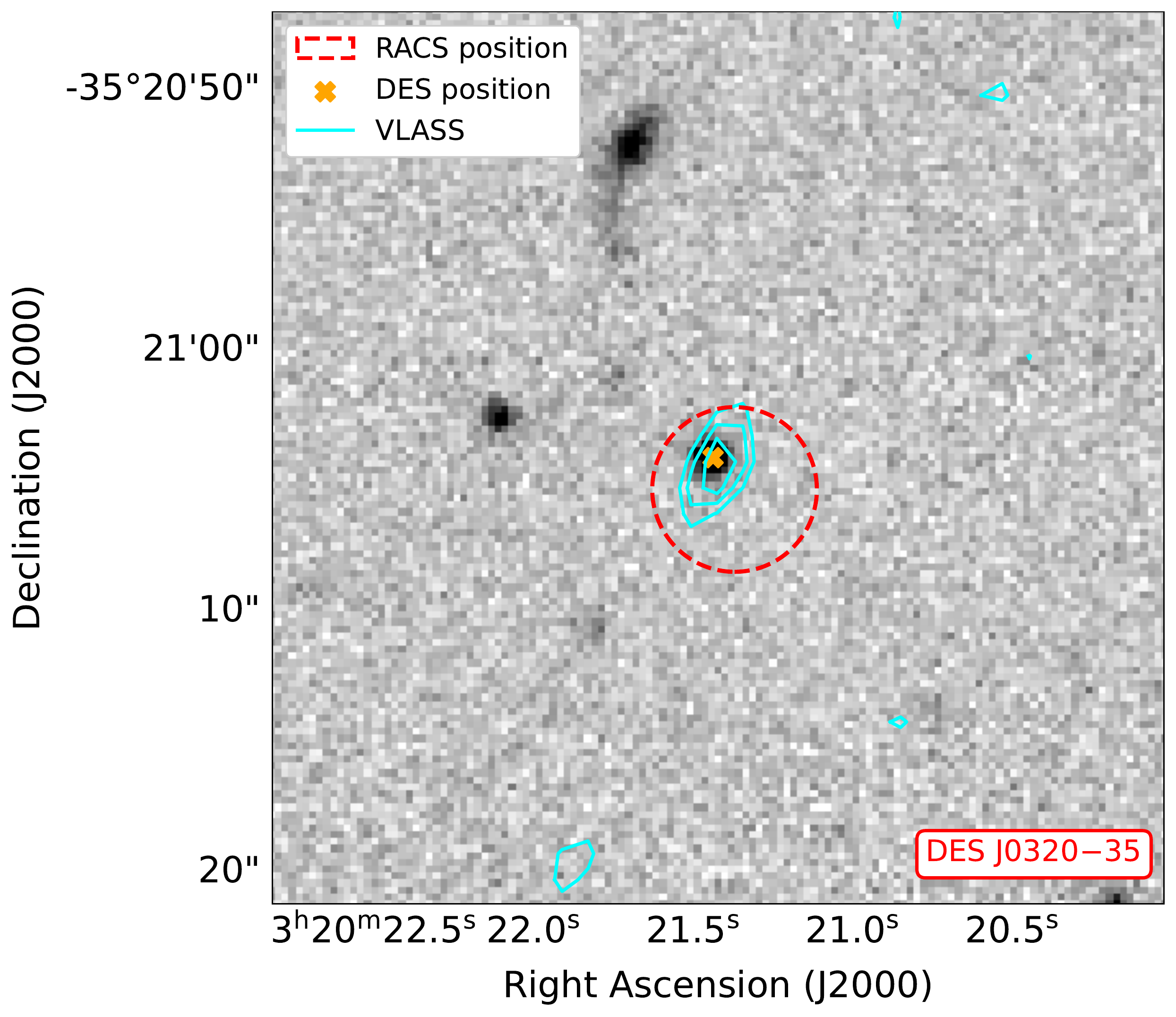}
	\includegraphics[width=0.625\hsize]{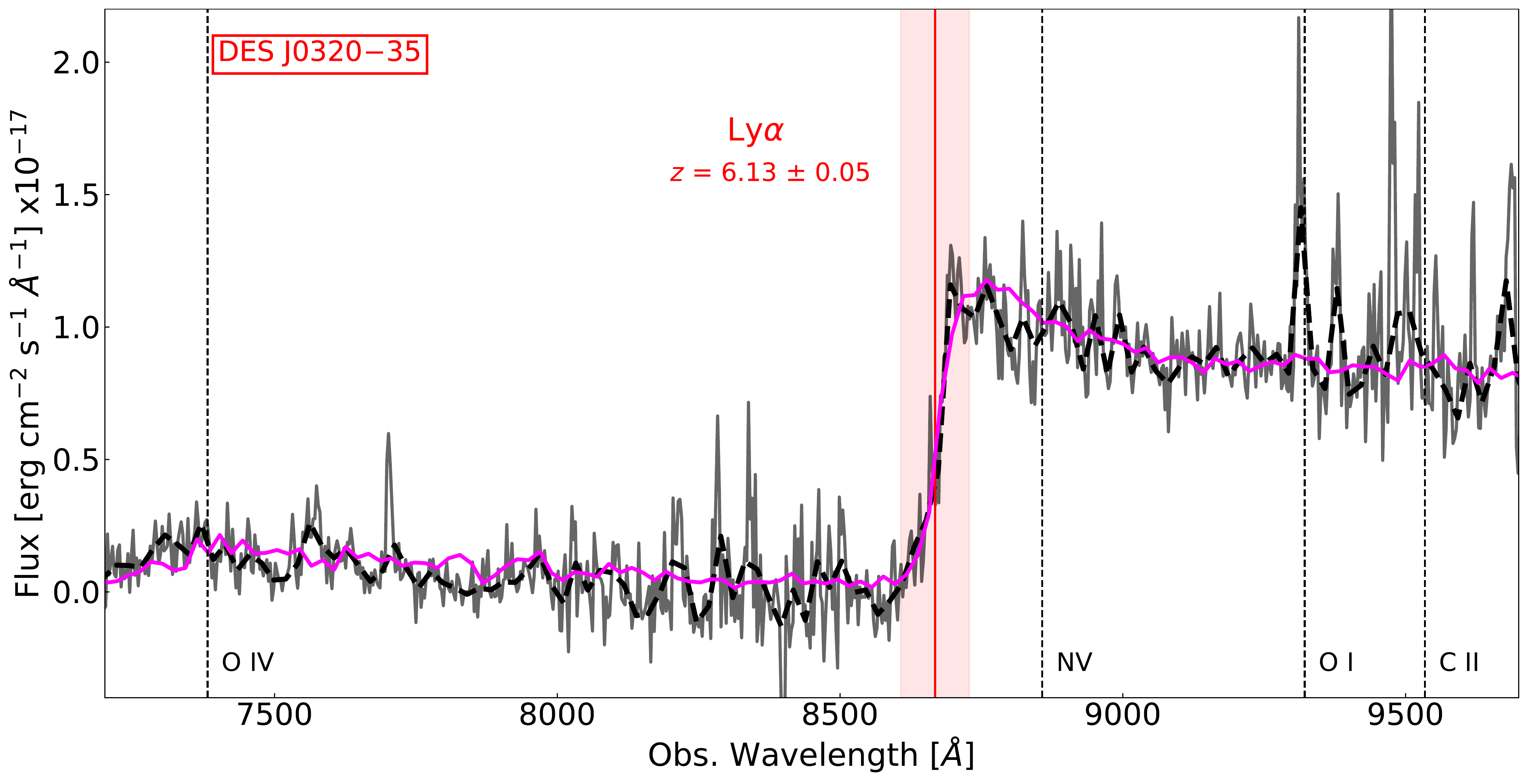}

    \includegraphics[width=0.37\hsize]{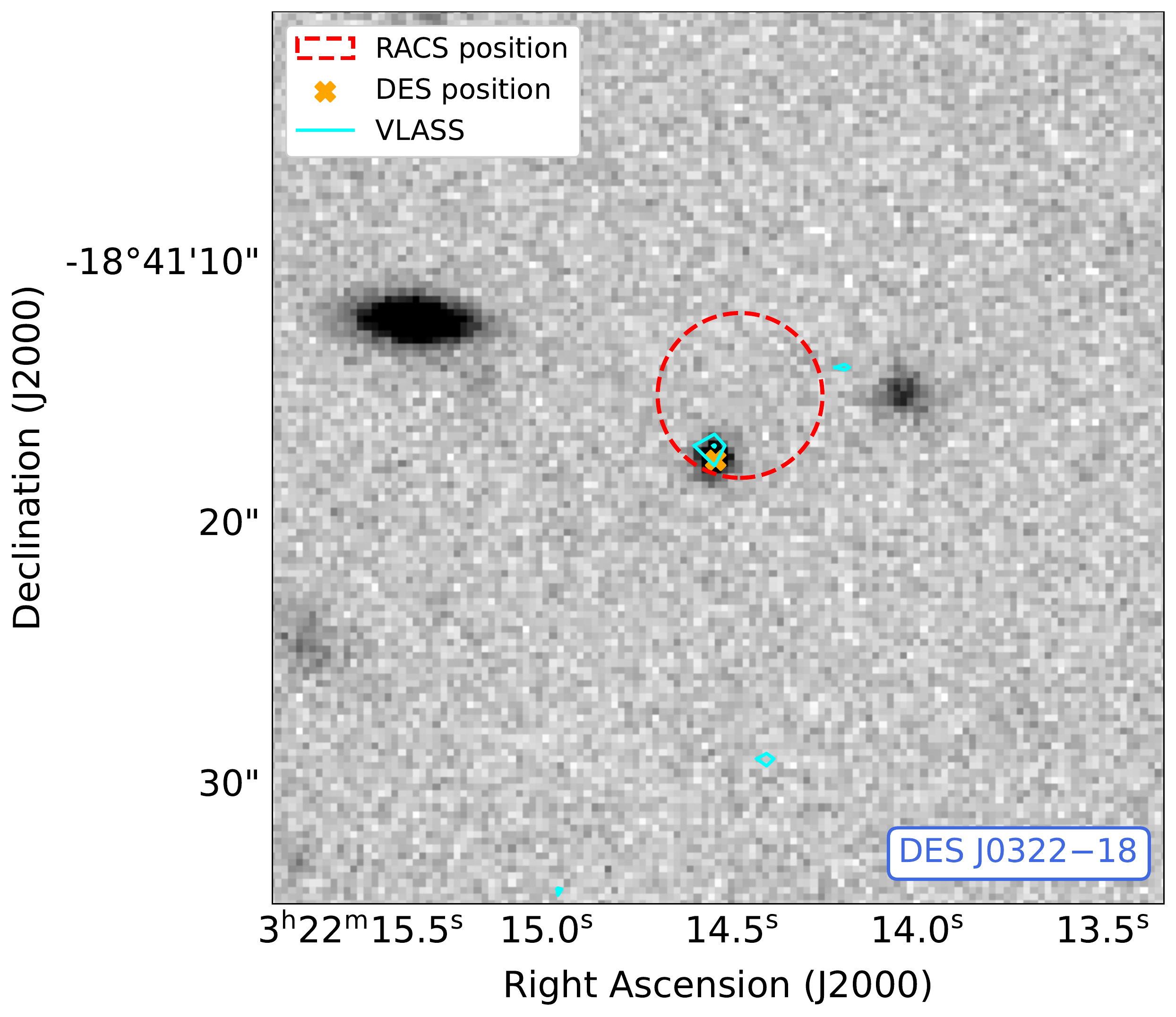}
	\includegraphics[width=0.625\hsize]{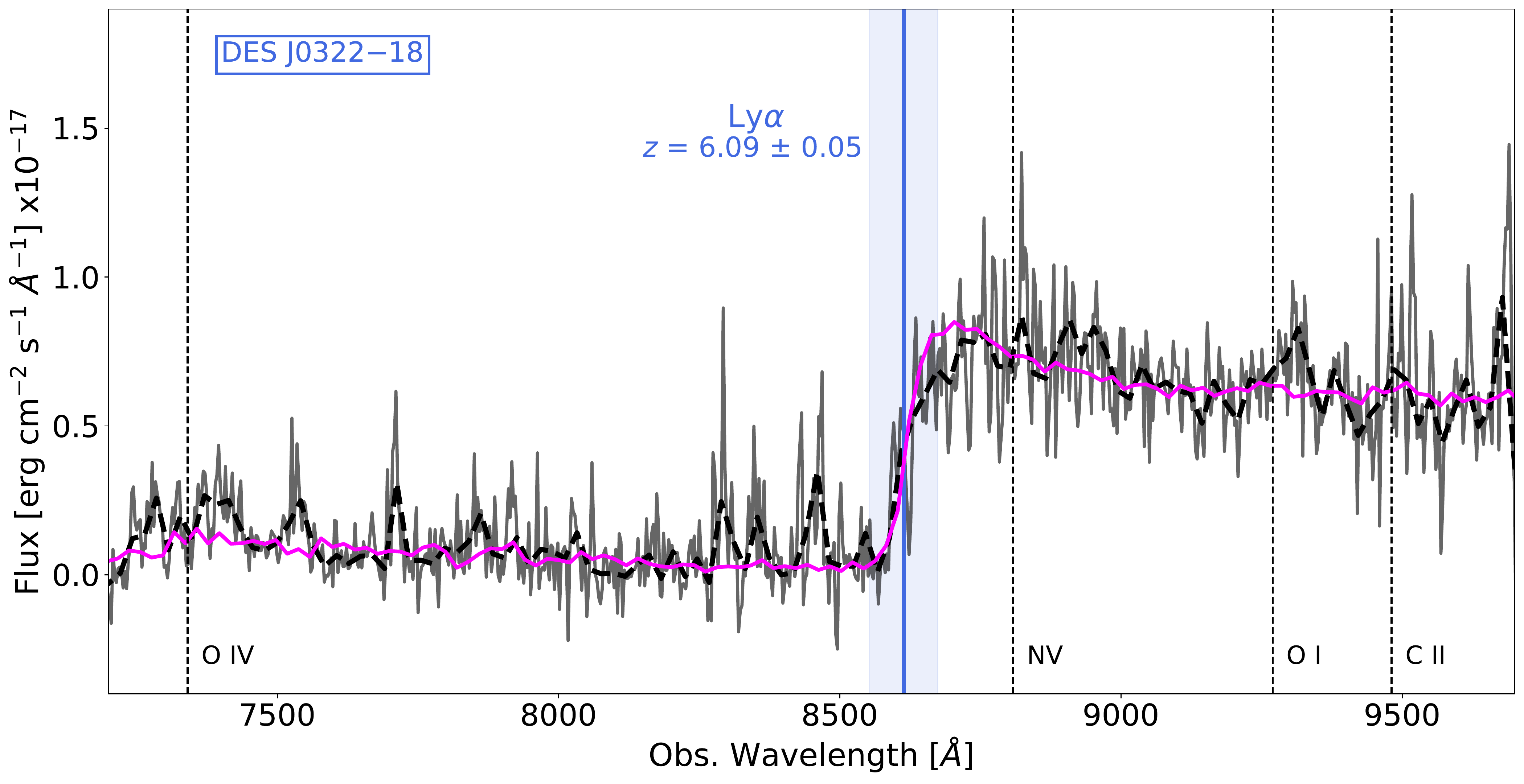}

\caption{{\bf Left:} 30\arcsec$\times$30\arcsec cutout around DES~J0320$-$35 (top) and DES~J0322$-$18 (bottom) in the $z_{\rm DES}$ filter. The optical and radio positions are shown with an orange cross and a dashed red circle (uncertainty reported in the RACS catalogue), respectively, while the VLASS radio (3~GHz) contours at 3,4,5 $\times$ RMS are displayed in cyan. {\bf Right:} Optical/NIR spectra of DES~J0320$-$35 (at $z=6.13\pm0.05$; top) and DES~J0318$-$18 (at $z=6.09\pm0.05$; bottom) obtained with the GMOS instrument on the Gemini-South telescope at  3\AA \, resolution. The dashed black lines are the spectra smoothed to a 20\AA \, resolution. The red and blue vertical lines indicate the expected wavelength of the Ly$\alpha$ emission line. The solid magenta line is the composite spectrum derived by {\protect\cite{Banados2016}} for $5.6<z<6.5$ QSOs with a weak Ly$\alpha$ emission line (like our sources), also smoothed to a 20\AA \, resolution.}
    \label{fig:opt_images}
\end{figure*}

While the spectroscopic observations of the majority of the sample are currently ongoing, we confirmed the high-$z$ nature of two candidates: DES~J0320$-$35 and DES~J0322$-$18. These two sources were selected as the first targets to be observed because they had the highest dropout in our sample ($i-z>3$, see Fig. \ref{fig:colour_cuts}) as well as a further radio association (\textless0.5\arcsec \, with S/N~$>3$) in the quick-look images of VLASS at 3~GHz.
In the left panels of Fig. \ref{fig:opt_images}, we report the optical (DES, orange cross) and radio positions (RACS, dashed red circle corresponding to the radio positional uncertainty of each source) as well as the VLASS contours overlaid on the optical images in the $z_\mathrm{DES}$ filter of both QSOs. 

\subsection{Spectroscopic confirmation and optical/NIR properties}
\label{Sec:spect_obs}

 We confirmed both candidates to be high-$z$ QSOs with the Gemini-South telescope, DES~J0320$-$35 on 2021 August 09 and DES~J0322$-$18 on 2021 October 14-15 (program ID: GS-2021A-DD-112; P.I. L. Ighina). The observations were carried out with the Gemini Multi-Object Spectrograph (GMOS; \citealt{Hook2004}) instrument in long-slit mode (1\arcsec \, aperture) and with the R400 grating. The mean airmass during the observations was 1.03 and 1.33, while the seeing was $\sim$0.9\arcsec \, and $\sim$0.7\arcsec \, for DES~J0320$-$35 and DES~J0322$-$18, respectively. Both targets were observed for a total of eight exposures, half with the central wavelength of the grism at 9000~\AA \, and half at 9100~\AA, in order to cover the spectral gap in the detector. For DES~J0320$-$35, each segment lasted 450~s (3600~s in total), while for DES~J0322$-$18, each segment lasted 800~s (6400~s in total). We also observed the spectro-photometric standard star LP~995$-$86 in order to correct the spectra of the two targets for the response of the instrument.

The data were then reduced using the dedicated IRAF Gemini package and following the instructions reported in the GMOS Data Reduction Cookbook (Version 1.1; Tucson, AZ: National Optical Astronomy Observatory; Shaw, R. A. 2016)\footnote{Available at: \url{http://ast.noao.edu/sites/default/files/GMOS\_Cookbook}.}. The RMS reached during the wavelength calibrations was $\sim$0.18\AA \, for both observations. Finally, since the nights of the observations were not in photometric conditions, in order to have an absolute flux calibration, the two spectra were normalised to the $z$ and $Y$ photometric data available from different surveys, by considering the magnitudes corrected by the Galaxy absorption as reported in the corresponding catalogue.

In both cases, the presence of a high drop in the continuum spectrum at about $\sim$8700\AA \, confirms the high-$z$ nature of the candidates. However, due to the lack of evident spectral lines in the observed range, we used the observed wavelength of the drop to estimate the best-fit redshift for both sources. In particular, we considered the composite spectrum derived by \cite{Banados2016} from a sample of 16 QSOs at $5.6<z<6.5$ with a weak Ly$\alpha$ emission line (rest-frame equivalent width, REW, $<15.4$\AA) and, after smoothing it to a 20\AA \, resolution, we used it to perform a fit to the observed spectrum, also smoothed to 20\AA, in the wavelength range 8000--9500\AA, where the only free parameters were the redshift and the normalisation. The best-fit redshift values we obtained are the following: $z=6.13\pm0.05$ for DES~J0320$-$35 and $z=6.09\pm0.05$ for DES~J0320$-$18. As error on these estimates we consider $3\times20\AA$ \, ($\sigma_{z}=\pm0.05$), which roughly corresponds to the width of the observed drops. Nevertheless, further observations with a larger wavelength range are needed to accurately determine their redshift.

The reduced spectra together with the high-$z$ QSOs composite spectra are reported in the right panels of Fig. \ref{fig:opt_images}. In Tab. \ref{tab:opt_data} we list the magnitudes of both QSOs (AB system\footnote{See \url{http://casu.ast.cam.ac.uk/surveys-projects/vista/technical/filter-set} and \url{https://wise2.ipac.caltech.edu/docs/release/allsky/expsup/sec4\_4h.html} for the conversion factors from Vega to AB system for the VHS/VIKING and WISE filters, respectively.}) in the optical and the IR available from the following surveys: DES, Pan-STARRS, the VISTA
Kilo-degree Infrared Galaxy Public Survey (VIKING; \citealt{Edge2013}), the VISTA Hemisphere Survey \citep[VHS;][]{McMahon2013} and the catWISE. In order to compute the optical spectral index of the two sources, we considered the power-law that best describes the continuum obtained from all the NIR photometric data points, that is, the filters $Y$, $J$, $H$, $K_s$, also from different surveys when available. The resulting spectral indices are as follows: $\alpha_\mathrm{o}^{\mathrm{\lambda}}=1.72\pm0.03$ ($\alpha_\mathrm{o}^{\mathrm{\nu}}=0.28$; DES~J0320$-$35) and $\alpha_\mathrm{o}^{\mathrm{\lambda}}=1.60\pm0.04$ ($\alpha_\mathrm{o}^{\mathrm{\nu}}=0.40$; DES~J0322$-$18), which are roughly consistent with the typical values found in QSOs \citep[e.g.][]{VandenBerk2001}. 

\begin{table}
	\centering
	\caption{Optical and IR magnitudes (AB system) of DES~J0320$-$35 and DES~J0322$-$18 available from the DES, VIKING, VHS and catWISE surveys. For DES~J0322$-$18 we report the magnitudes from PanSTARRS in brackets. Upper limits are given at a 5$\sigma$ significance level. We also report the slope of the continuum and the rest-frame luminosities at 2500 and 4400~\AA.}
	\label{tab:opt_data}
	\begin{tabular}{lccccc} 
		\hline
		Filter & DES~J0320$-$35 & DES~J0322$-$18\\
		\hline
		$r_\mathrm{{DES}}$ \: ($r_\mathrm{{PS}}$) & \textgreater24.56 & \textgreater24.56 \: (\textgreater23.20) \\
		$i_\mathrm{{DES}}$ \: ($i_\mathrm{{PS}}$) & 23.70$\pm$0.24 & \textgreater23.96 \: (\textgreater23.10)\\
		$z_\mathrm{{DES}}$ \: ($z_\mathrm{{PS}}$) & 20.59$\pm$0.03 & 20.94$\pm$0.04 \: (21.77$\pm$0.07)\\
		$Y_\mathrm{{DES}}$ \: ($Y_\mathrm{{PS}}$) & 20.36$\pm$0.08 & 20.86$\pm$0.11\: (20.77$\pm$0.12)\\
		$z_\mathrm{{Vista}}$ & 20.88$\pm$0.05 & --\\
		$Y_\mathrm{{Vista}}$ & 20.33$\pm$0.07 & --\\
		$J_\mathrm{{Vista}}$ & 20.24$\pm$0.08 & 20.55$\pm$0.12\\
		$H_\mathrm{{Vista}}$ & 20.20$\pm$0.11 & --\\
		$K_\mathrm{{Vista}}$ & 20.11$\pm$0.17 & 20.68$\pm$0.39\\
		$W_\mathrm{{1}}$   & 20.01$\pm$0.06 & 20.28$\pm$0.08\\
		$W_\mathrm{{2}}$   & 20.27$\pm$0.12 & 20.84$\pm$0.23\\
		\hline
		$\alpha_\mathrm{o}^\mathrm{\lambda}$ & 1.72$\pm$0.03 & 1.60$\pm$0.04\\
		L$_\mathrm{2500\AA}$ (erg~s$^{-1}$~Hz$^{-1}$) & 1.80$^{+0.21}_{-0.19} \, \times 10^{31}$ & 1.20$^{+0.19}_{-0.16} \, \times 10^{31}$\\
		L$_\mathrm{4400\AA}$ (erg~s$^{-1}$~Hz$^{-1}$) & 2.11$^{+0.24}_{-0.22} \, \times 10^{31}$ & 1.52$^{+0.24}_{-0.21} \, \times 10^{31}$ \\

		\hline
	\end{tabular}
\end{table}
\subsection{Archival Radio Data}

To characterise the radio properties of these two newly discovered QSOs, we searched for archival data from past and current radio surveys.
As already mentioned, both QSOs were first selected from RACS, where they were detected with a high S/N  (\textgreater 11 for DES~J0320$-$35, off-source RMS$\sim$0.28~mJy~beam$^{-1}$, and \textgreater8 for DES~J0322$-$18, off-source RMS$\sim$0.19~mJy~beam$^{-1}$, according to the catalogue derived by \citealt{Hale2021}). Furthermore, their radio association was also confirmed by the VLASS quick-look images, which have an higher angular resolution ($\sim$2.5\arcsec). As the VLASS quick-look images might be unreliable at low flux densities, we added a further 10 per cent to the uncertainty of their flux densities in this work (for details, see \citealt{Gordon2021}).

At the same time, we also checked the other publicly available radio surveys. At low frequency, we did not find a counterpart in the TIFR Giant Metrewave Radio Telescope Sky Survey \citep[TGSS;][]{Intema2017}, at 148~MHz, nor in the GaLactic and Extragalactic All-sky Murchison Widefield Array South Galactic Pole \citep[GLEAM-SGP;][]{Franzen2021}, at 216~MHz\footnote{Here we only consider the observations in the wide-band 200--231~MHz image, which provide the strongest constraints.}. The RMS values between our two sources differ significantly  due to the presence of a very bright object near DES~J0320$-$35.
Moreover, DES~J0322$-$18 is also detected in the NRAO VLA Sky Survey \citep[NVSS;][]{Condon1998} at 1.4~GHz with a 4$\sigma$ significance (RMS$\sim$0.37~mJy~beam$^{-1}$, at a distance \textless5\arcsec), even if not reported in the catalogue. Near DES~J0320$-$35 a radio signal is also present, but the corresponding emission peak is 25\arcsec \, away. Since this distance is larger than the pixel size of the NVSS images (15\arcsec) we do not consider the corresponding flux density. 

\begin{figure}
	\includegraphics[width=\hsize]{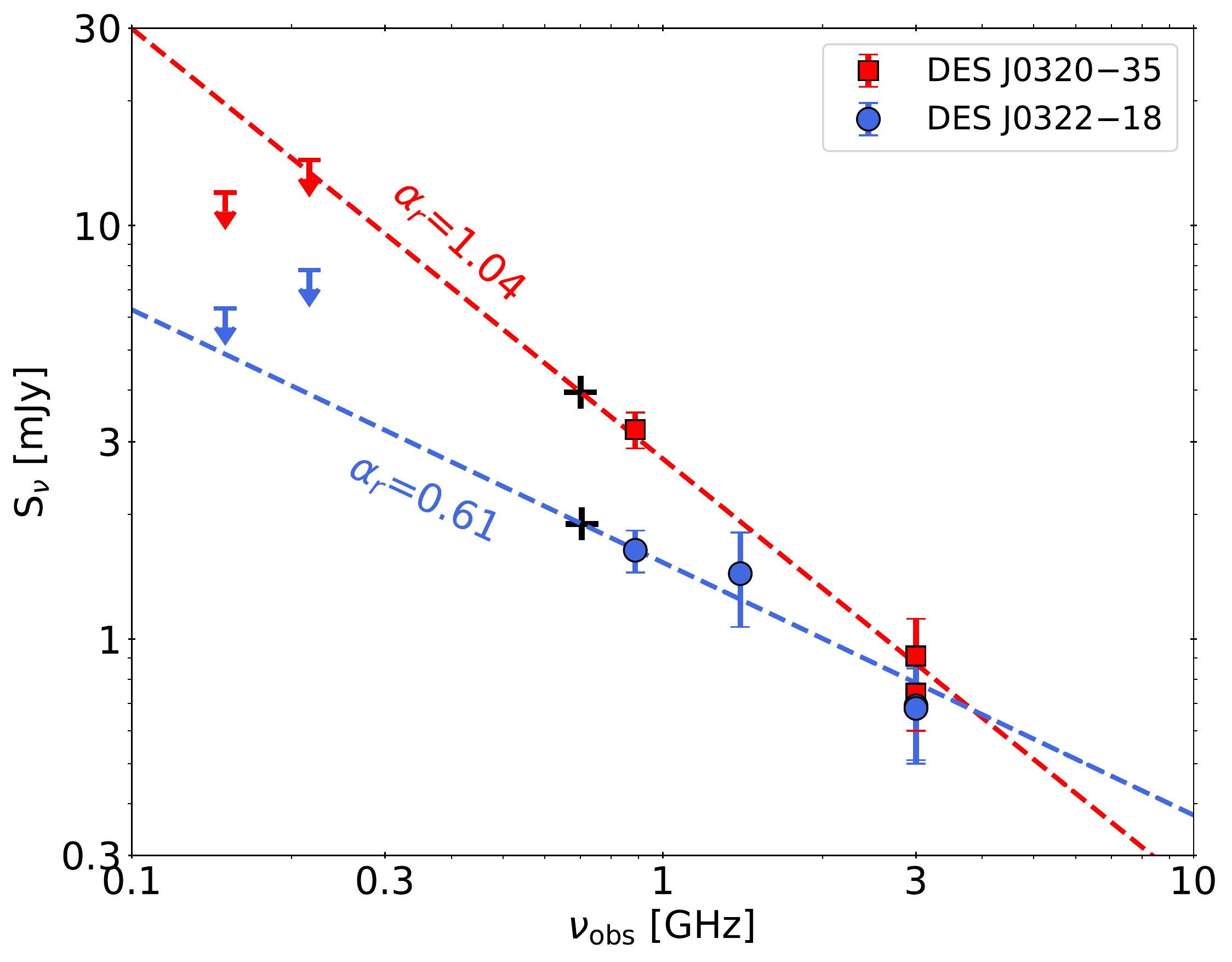}
    \caption{Radio data available for DES~J0320$-$35 (red squares) and DES~J0322$-$18 (blue circles). The dashed lines represent the single power law that best fit the data of each source. The black crosses correspond to the extrapolated flux density at the 5~GHz rest-frame frequency of each object.}
    \label{fig:radio_spectra}
\end{figure}

We summarise in Tab. \ref{tab:radio_data} the flux density measurements of both sources in the radio surveys described above. For both QSOs we consider the peak flux density of each image since neither of them is resolved. In case of a non-detection, we report the upper limit computed as 3$\times$local RMS.
From both the detections and non-detections, the power-law that best describe the two radio spectra have spectral indices of $\alpha_\mathrm{r}=1.04\pm0.07$ (DES~J0320$-$35) and $\alpha_\mathrm{r}=0.61\pm0.11$ (DES~J0322$-$18), obtained with the \texttt{MRMOOSE} code \citep{Drouart2018b,Drouart2018}. In Fig. \ref{fig:radio_spectra} we show the radio detections and upper limits of both QSOs together with the best-fit power-law.
These values are consistent with the ones typically observed in RL AGNs \citep[e.g.][]{Banados2021}, even though they are relatively uncertain given the few data available and the potential intrinsic variability. From the combination of the LoTSS (144~MHz) and the Faint Images of the Radio Sky at Twenty-Centimeters (FIRST, 1.4~GHz; \citealt{Becker1995}) surveys, \cite{Gloudemans2021,Gloudemans2022} found that $z>5$ QSOs have a median spectral index of  $\alpha_{\rm r}=0.3$. This difference can be attributed to two main factors: the frequency range probed by the different works and the unprecedented sensitivity reached by LoTSS. Indeed, while typical high-$z$ RL QSOs have spectral indices of $\sim0.7-1$ \citep[e.g.][]{Banados2015,Spingola2020}, many of them also present a turnover in the radio spectrum at observed frequencies of 0.5--2~GHz \citep[e.g.][]{Shao2022,Ighina2022b, Belladitta2022b}, which can be interpreted as a nearly flat spectrum from a two-point spectral index with a measurement at very low frequencies. Moreover, since the FIRST survey is shallower, compered to LoTSS, the sources with a measurement in both surveys are likely to have a nearly flat or even negative spectral index (see sec. 3 in \citealt{Gloudemans2021}).

Interestingly, the non-detection in the TGSS survey at 148~MHz might imply a flattening of the radio spectrum or the presence of a turnover at lower frequencies also for DES~J0320$-$35. While for DES~J0322$-$18, given its relatively flat spectrum, the upper limits at low frequencies are still consistent with the single power law extrapolation. In both cases more radio observations are needed to accurately constrain the spectrum and the nature of the two RL QSOs.


\begin{table}
	\centering
	\caption{Radio data available from public surveys for DES~J0320$-$35 and DES~J0322$-$18. Since the sources are not resolved, we report the peak flux density measured in each image. Upper limits are at a 3$\sigma$ significance level. We also report the best-fit spectral index, the luminosity density at 5~GHz rest frame and the radio-loudness parameter.}
	\label{tab:radio_data}
	\begin{tabular}{lccccc}
		& & \multicolumn{2}{c}{DES~J0320$-$35} &  \multicolumn{2}{c}{DES~J0322$-$18} \\
		\hline
		Survey & Frequency & $S_\nu$ & RMS & $S_\nu$ & RMS\\
		& (MHz) & \multicolumn{2}{c}{\: (mJy beam$^{-1}$)}& \multicolumn{2}{c}{\: (mJy~beam$^{-1}$)} \\
		\hline
		TGSS & 148 & \textless12.2 & 4.1 & \textless6.3 & 2.1\\
		GLEAM-SGP & 216 & \textless14.5 & 4.8 & \textless7.8 & 2.6\\
		RACS & 888 & 3.21$\pm$0.28 & 0.28 & 1.64$\pm$0.19 & 0.19\\
		NVSS & 1400 & -- & 0.45 & 1.44$\pm$0.37 & 0.37\\
		VLASS (1) & 3000 & 0.74$\pm$0.14 & 0.12 & 0.69$\pm$0.19 & 0.17\\
		VLASS (2) & 3000 & 0.91$\pm$0.18 & 0.16 & 0.68$\pm$0.17 & 0.15\\
		\hline
		$\alpha_\mathrm{r}$ & & \multicolumn{2}{c}{1.04$\pm$0.07} &  \multicolumn{2}{c}{0.61$\pm$0.11} \\
		L$_\mathrm{5GHz}$ (W Hz$^{-1}$) &  &\multicolumn{2}{c}{2.42$ ^{+0.25}_{-0.25}\times$10$^{26}$}  &  \multicolumn{2}{c}{1.10$^{+0.16}_{-0.15} \, \times$10$^{26}$} \\
		$R$=$S_{\rm 5 GHz}$/${S_{\rm 4400 \AA}}$& & \multicolumn{2}{c}{115$\pm$18} &  \multicolumn{2}{c}{73$\pm$15} \\
		\hline
		
	\end{tabular}
\end{table}

\subsection{Lack of spectral emission lines}

Based on the optical/NIR spectra obtained from Gemini-GMOS, neither source shows prominent emission lines in the observed range (namely Ly$\alpha$). Given their RL nature, the lack of emission lines could be associated with a BL Lacertae (BL Lac) nature \citep[REW $<5\AA$; e.g.][]{Padovani2017}. If confirmed, these would be the farthest BL Lac objects currently known, with other three potential candidates at $z>6$ \citep[see][]{Koptelova2022,Gloudemans2022}. In this case, we would expect to observe a nearly flat radio spectrum; simultaneous observations in the radio band would be needed to reliably constrain the spectral slopes of these sources.
Another possibility could be that these are weak emission-line QSOs (WLQSOs) that happen to host two powerful relativistic jets (similarly to the $z=6.18$ RL QSO J1429+5447, \citealt{Shen2019},  and the ones discussed in \citealt{Gloudemans2022}). As their name suggests, these are QSOs with very weak emission lines (REW[CIV1549\AA] $<15.4$\AA; \citealt{Diamond-Stanic2009}). The physical conditions of these sources are not fully clear. The two main scenarios proposed to explain the lack of strong emission lines involve either a young accretion system \citep[e.g.][]{Eilers2018,Andika2020,Andika2022}, where the Broad Line Region (BLR) has not fully formed yet \citep[e.g.][]{Meusinger2014}, or a softer ionising continuum, due to different accretion properties or absorption between the disc and the BLR \citep[e.g.][]{Luo2015}.

Interestingly, recent works did find an increase of the WLQSO fraction at high redshift being $\sim$10-14\% at $z\gtrsim5.7$ (e.g. \citealt{Banados2016, Shen2019}), as opposed to $\sim$6\% found by \cite{Diamond-Stanic2009} considering $3<z<5$ QSOs. For the RL population the observed evolution is even more pronounced, with $\sim$38\% RL QSOs at $z>5$ being classified as WLQSOs \citep[see e.g.][]{Gloudemans2022}. Therefore, this redshift evolution can indicate possible changes in the accretion mode of SMBHs and also in the RQ/RL populations. However, current estimates from $z>5$ studies are still limited by small number statistics.
If the weak-line nature of both objects was confirmed, it would further strengthen the presence of an increase of WLQSOs as a function of redshift. However the spectroscopic identification of the remaining candidates selected in this work is needed to have an estimate of the weak-line fraction in our sample. 
Moreover, we stress once again that a NIR spectrum covering a larger wavelength range and with a higher S/N is needed in order to understand conditions of the accretion and of the BLR in these two confirmed high-$z$ QSOs. 

\section{Comparison with expectations}
\label{sec:expectations}

Thanks to its wide area coverage and relatively deep radio/optical sensitivity, the combination of RACS+DES is currently one of the most effective tools for selecting high-$z$ RL QSOs. From our work, we can derive an estimate of the space density of RL QSOs with $S_{\rm 888MHz}>1$~mJy~beam$^{-1}$ and ${\rm mag}(z_\mathrm{{DES}})<21.3$ at $5.9<z<6.4$, where our selection is nearly complete ($\sim$71 per cent, considering both the optical/NIR and radio completeness, see Sec. \ref{sec:selection}). Based on the discovery of DES~J0320$-$35 and DES~J0322$-$18 over an area of 5000~deg$^2$ (corresponding to a comoving volume slice of 21.5~Gpc$^3$ between $5.9<z<6.4$) the resulting space density is 0.13$^{+0.18}_{- 0.09}$~Gpc$^{-3}$. As uncertainties we considered the Poissonian error (lower) and the total number of candidates selected from RACS (5; upper). However, a comparison with the expectations from lower redshift is not straightforward. This is because the RL population is composed by a large variety of sources, whose observed properties can be affected by relativistic beaming at different degrees (depending, for example, on the orientation of the relativistic jets) and therefore modify the observed shape of their luminosity function (LF; e.g. \citealt{Urry1995}). For this reason, the single classes of RL AGNs are often studied separately (e.g. \citealt{Rigby2011, Diana2022}).
However, since it is not possible to exclude that some degree of relativistic boosting is present in DES~J0320$-$35 and DES~J0322$-$18 with the data currently available, we estimated the overall number of RL QSOs expected to be found in a given survey using the following approach.

We started from the radio LF derived by \cite{Mao2017} for the flat-spectrum radio quasar (blazar hereafter) population, that is, sources with the relativistic jet oriented close to our line of sight ($\theta_{\rm view}<1/\Gamma$, where $\Gamma$ is the bulk Lorentz factor of the jet) and, therefore, for which we expect the radio spectrum to be dominated by the radiation relativistically boosted. The LF derived by \cite{Mao2017} was built using blazars in the redshift range $z=0.5-3$ and then was confirmed up to $z\sim5$ by \cite{Caccianiga2019}. From this LF we created a mock sample of blazars at $z>5$ whose optical luminosities were computed based on the optical-to-radio ratio distribution observed in the sample discussed by \cite{Caccianiga2019}, which corresponds to the highest-redshift complete sample of blazars currently available.
In this way we built a sample of blazars for which the redshift, the radio flux density and the optical/NIR magnitude are known (we assumed $\alpha_{\rm r}=0$, \citealt{Padovani2017}, and $\alpha_{\rm o}=0.44$, \citealt{VandenBerk2001}). By considering different radio/optical limits, we were then able to determine the corresponding number of detectable blazars for a given radio/optical survey combination.
Finally, in order to estimate the overall number of RL QSOs starting from blazars, we computed the number of RL QSOs with a misaligned jets whose radio flux density is relativistically de-beamed\footnote{The (de-)amplification factor scales as $\propto \delta^{p}$, where the Doppler factor, $\delta$, is given by $[\Gamma(1-\beta {\rm cos}\theta_{\rm view})]^{-1}$ and $p=2+\alpha_{\rm r}$ \citep[e.g.][]{Cohen2007}.}, but still detectable in the given radio survey, following the approach outlined in section 2 of \citealt{Ghisellini2016}.  As a reference, this last step resulted in an increase of a factor $\sim$10 in the number of total RL QSOs (with respect to blazars only) for $S_{\rm 888MHz}>1$~mJy~beam$^{-1}$ and $5.9<z<6.4$.

\begin{figure}
\centering
	\includegraphics[width=\hsize]{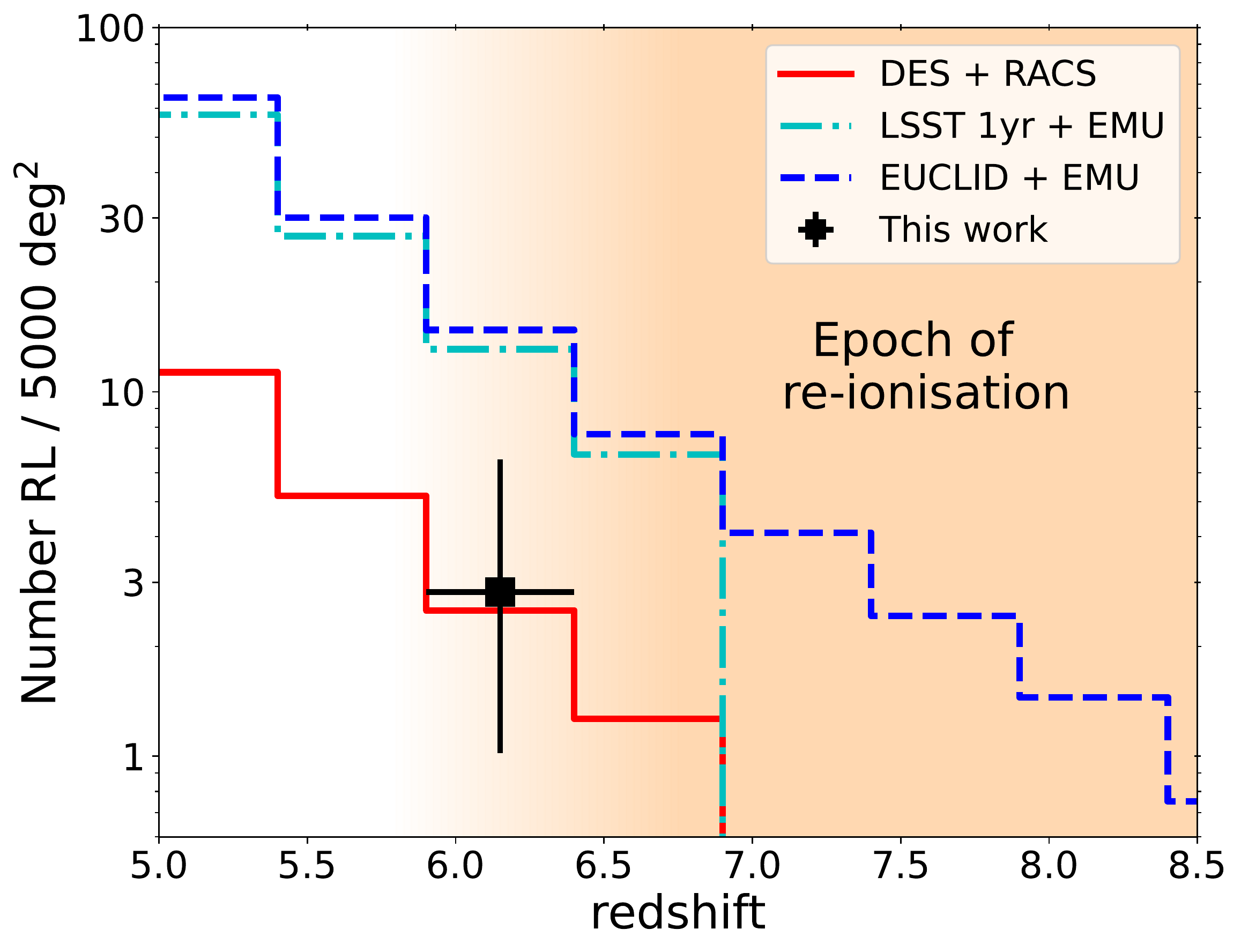}
    \caption{Expected number of $z>5$ RL QSOs per redshift bin in different combinations of current/future radio+optical/NIR surveys rescaled to an area of 5000~deg$^2$. The sensitivities adopted for the different surveys are the following: DES mag($z$-band) = 21.3, VRO/LSST 1yr mag($z$-band) = 23, \textit{Euclid}--WIDE mag(NIR)=24,} RACS $S_{\rm 888MHz}$ = 1~mJy~beam$^{-1}$ and EMU $S_{\rm 944MHz}$ = 0.1~mJy~beam$^{-1}$. In the redshift bin $5.9<z<6.4$ we also show the constraints derived from the discovery of the two new RL QSOs from the RACS+DES combination discussed in the work. The estimate has been corrected for the completeness of the selection described in Sec. \ref{sec:selection}, while the upper error is given by the completeness-corrected number of candidates we selected from RACS.
    \label{fig:predictioons}
\end{figure}

In Fig. \ref{fig:predictioons}, we show the expected number of RL QSOs per bin of redshifts to be detectable in different combinations of radio+optical/NIR surveys. In particular, in the radio band we considered RACS (limit 1~mJy~beam$^{-1}$; \citealt{Hale2021}) and the EMU survey (limit 0.1~mJy~beam$^{-1}$; \citealt{Norris2021}), while in the optical band we considered DES (limit mag($z$-band) = 21.3, as in this work), the Vera C. Rubin Legacy Survey of Space and Time (VRO/LSST; limit mag($z$-band) = 23 after the first year; \citealt{Ivezic2019}) and the {\it Euclid} wide survey (limit mag(NIR) = 24; \citealt{Euclid2022}). Moreover, in the case of DES and VRO/LSST, we only show the expected number of sources up to $z=6.9$, since the wavelength coverage of these surveys will not allow for the detection of objects at higher redshift. 


Based on the extrapolations from lower redshift, the number of expected RL QSOs with S$_{\rm 888MHz}>1$~mJy~beam$^{-1}$ and ${\rm mag}(z_\mathrm{{DES}})<21.3$ at $5.9<z<6.4$ is about three. This is broadly consistent with the discovery of two new RL QSOs at $\sim6.1$. Indeed, by taking the completeness of our selection into account ($\sim71$ per cent , see Sec. \ref{sec:selection}), the expected number of sources is $2.8_{-1.8}^{+3.7}$ (black square in Fig. \ref{fig:predictioons}), where we computed the upper error by considering the total number of high-$z$ candidates found in this work from RACS.

Based on Fig. \ref{sec:expectations}, if we consider instead the upcoming EMU and VRO/LSST surveys, we expect to increase the number of RL QSOs up to $z\sim6.5$ by more than an order of magnitude. This is due both to the deeper sensitivities, in the radio and in the optical/NIR bands, as well as to the much larger common area (the entire southern sky; \citealt{Ivezic2019,Norris2013}). However, the VRO/LSST, similarly to DES, will only be sensitive to sources up to $z\sim7$, after which we expect all the optical emission of QSOs to be redshifted at wavelengths outside the $grizY$ filters. At the same time, other complementary surveys such as the {\it Euclid} wide survey, with filters also in the NIR ($YJH$), will reach a similar magnitude limit at higher wavelengths (e.g. \citealt{Euclid2019}) and therefore, together with the EMU survey, they will be able to explore the RL QSO population up to $z\sim8.5$.

\section{Summary and Conclusions}
\label{Sec:summary}
We have presented the selection of a sample of high-$z$ RL QSOs in the southern hemisphere based on their optical/NIR colours as reported in DES and their detection in the RACS radio survey. We set the selection criteria such that we were able to recover almost all previously discovered QSOs in the same area above our optical/radio flux limits.
The completeness and the efficiency of our selection, mainly driven by the radio catalogue for the association, will significantly increase with the upcoming RACS data releases.

From the sources selected in this work, we were already able to identify two new high-$z$ RL QSOs: DES~J0320$-$35 at $z=6.13\pm0.05$ and DES~J0322$-$18 at $z=6.09\pm0.05$. These are now two of the most distant RL QSOs currently known. Indeed only a few $z>6$ RL QSOs have been detected in the radio band \citep[e.g.][]{Liu2021, Gloudemans2021} and even fewer are confirmed as radio loud \citep[e.g.][]{Banados2021, Gloudemans2022}. Interestingly, both sources do not present prominent broad emission lines, suggesting that they are both WLQSOs. 
If confirmed by NIR further spectroscopy, this would further strengthen the increase of the WLQSOs fraction at high redshift recently hinted both in the
RQ \citep[e.g.][]{Banados2016,Shen2019} and the RL \citep[e.g.][]{Gloudemans2022} QSO populations and therefore imply an evolution of SMBH accretion conditions/modes at high redshift.

By comparing the number of selected and confirmed candidates to the expected number of RL QSOs at $5.9<z<6.4$ with the same flux density cuts ($S_{\rm 888MHz}>1$~mJy~beam$^{-1}$ and ${\rm mag}(z_\mathrm{{DES}})<21.3$), we found that the discovery of DES~J0320$-$35 and DES~J0322$-$18 is consistent with expectations. 
Moreover, we also showed how the upcoming wide-area surveys (e.g. VRO/LSST, {\it Euclid} wide and EMU) can increase the number of RL QSOs at $z>5$ by a factor $> 10$ and potentially reach sources up to $z\sim8.5$. Combined with samples of high-$z$ radio galaxies selected with different techniques \citep[e.g.][]{Saxena2018a,Saxena2018b,Saxena2019,Drouart2020,Broderick2022}, these sources can be used to address important questions of modern astrophysics: on the formation and growth of SMBHs \citep[e.g.][]{Overzier2022}, the evolution of relativistic jets properties \citep[e.g.][]{Ighina2022a}, their impact on the environment \citep[e.g.][]{Hardcastle2020} and the properties of the IGM in the early Universe \citep[e.g.][]{Carilli2004,Furnaletto2006}.

\section*{Acknowledgements}
 We thank the referee for their suggestions that have improved the quality of the paper. This work is based on observations obtained at the international Gemini Observatory, a program of NSF’s NOIRLab, which is managed by the Association of Universities for Research in Astronomy (AURA) under a cooperative agreement with the National Science Foundation. on behalf of the Gemini Observatory partnership: the National Science Foundation (United States), National Research Council (Canada), Agencia Nacional de Investigaci\'{o}n y Desarrollo (Chile), Ministerio de Ciencia, Tecnolog\'{i}a e Innovaci\'{o}n (Argentina), Minist\'{e}rio da Ci\^{e}ncia, Tecnologia, Inova\c{c}\~{o}es e Comunica\c{c}\~{o}es (Brazil), and Korea Astronomy and Space Science Institute (Republic of Korea).\\
In this work we made use of the Gemini IRAF package. IRAF is distributed by the National Optical Astronomy Observatory, which is operated by the Association of Universities for Research in Astronomy (AURA) under a cooperative agreement with the National Science Foundation.\\
We acknowledge financial contribution from the agreement ASI-INAF n. I/037/12/0 and n.2017-14-H.0 and from INAF under PRIN SKA/CTA FORECaST. We acknowledge financial support from INAF under the project `QSO jets in the early Universe', Ricerca Fondamentale 2022.\\
This research made use of Astropy (\url{http://www.astropy.org}) a community-developed core Python package for Astronomy \citep{astropy2018}. This research has made use of the VizieR catalogue access tool, CDS, Strasbourg, France (DOI : 10.26093/cds/vizier). The original description of the VizieR service was published in 2000, A\&AS 143, 23.

\section*{Data Availability}
All of the data used in this work are publicly available as described in the text. The DES catalogue can be retrieved from the corresponding SQL portal (\url{https://des.ncsa.illinois.edu/desaccess/home}), while the RACS source lists used for the radio association can be found on the CASDA portal under the AS110 project code (\url{https://data.csiro.au/domain/casdaObservation?redirected=true}). Reprocessed data are also available upon reasonable request to the corresponding author.



\bibliographystyle{mnras}
\bibliography{main} 





\bsp	
\label{lastpage}
\end{document}